\pdfoutput=1
\documentclass[twocolumn,tighten,apj,iop]{emulateapj}
\usepackage{amsmath, amsthm, amssymb}
\usepackage{graphics}

\newcommand{\sci}[2]{#1\times 10^{#2}}

\newcommand{\be}{\begin{equation}}
\newcommand{\ee}{\end{equation}}
\newcommand{\bea}{\begin{eqnarray}}
\newcommand{\eea}{\end{eqnarray}}
\newcommand{\ba}{\begin{array}}
\newcommand{\ea}{\end{array}}

\long\def\symbolfootnote[#1]#2{\begingroup%
\def\thefootnote{\fnsymbol{footnote}}\footnote[#1]{#2}\endgroup}

\shorttitle{Electromagnetic Transients from NSNS/BHNS Mergers}
\shortauthors{Roberts et al.}

\begin{document}
\title{Electromagnetic Transients Powered by Nuclear Decay in the Tidal Tails of Coalescing 
Compact Binaries}

\author{L. F. Roberts\altaffilmark{1},  D. Kasen\altaffilmark{2,3}, W. H. Lee\altaffilmark{4}, 
and E. Ramirez-Ruiz\altaffilmark{1}}
\altaffiltext{1}{Department of Astronomy and Astrophysics, University
of California, Santa Cruz, CA 95064 USA}
\altaffiltext{2}{Departments of Physics and Astronomy, University of California, Berkeley, 94270, USA}
\altaffiltext{3}{Nuclear Science Division, Lawrence Berkeley National Laboratory,  Berkeley, CA, 94720, USA}
\altaffiltext{4}{Instituto de Astronom\'{\i}a, Universidad Nacional Aut\'{o}noma 
de M\'{e}xico, Apdo. Postal 70-264, Cd. Universitaria, M\'{e}xico DF 04510}

\begin{abstract}
The possibility that long tidal tails formed during compact object mergers 
may power optical transients through the decay of freshly synthesized $r$-process 
material is investigated. Precise modeling of the merger dynamics allows for 
a realistic determination of the thermodynamic conditions in the ejected debris. 
The results of hydrodynamic and full nuclear network calculations are combined 
to calculate the resultant $r$-process abundances and the heating of the material 
by their decays.  The subsequent homologous structure is mapped into a radiative 
transfer code to synthesize emergent model light curves and determine  how their 
properties (variability and color evolution) depend on the mass ratio and orientation 
of the merging binary.  The radiation emanating from the ejected debris, though less 
spectacular than a typical supernova, should be observable in transient surveys and we 
estimate the associated  detection rates.  The case for (or against) compact object mergers 
as the progenitors of short gamma-ray bursts can be tested if such electromagnetic 
transients are detected (or not) in coincidence with some bursts, although they 
may be obscured by on-axis afterglows. 
\end{abstract} 

\keywords{nuclear reactions, nucleosynthesis, abundances --- black hole physics --- radiative transfer --- stars: neutron --- hydrodynamics --- gamma-ray burst: general} 
\maketitle 

\section{Introduction}
Merging compact binaries are the primary candidate for direct detection 
of gravitational waves (GWs) by LIGO \citep{Abbott08}, and are thought to be 
the progenitors of short gamma-ray 
bursts (SGRBs)\citep[c.f.][]{Lee07,Rosswog07,Nakar07,Gehrels09}.  Current 
observational limits indicate that any supernova-like event accompanying 
SGRBs would have to be over 
50 times fainter than normal Type Ia SNe or Type Ic supernovae (SNe), 
\citep{Hjorth05,Bloom06}.  These limits strongly constrain progenitor 
models for SGRBs \citep{Lee07,Perley09,Kocevski10}.  Unless SGRBs are 
eventually found to be accompanied by telltale optical signatures like 
the supernovae of long-duration GRBs \citep{Hjorth03}, the only 
definitive understanding of the progenitors will come from possible 
associations to direct gravitational or neutrino signals \citep{Rosswog03a}.

The merger of compact objects does not 
necessarily imply the absence of optical or other long-wavelength
phenomena after coalescence.  Neutron-rich material may be dynamically
ejected during a NS-NS (Neutron Star) or a NS-BH (Black Hole) merger.
Material dynamically stripped from a star is violently ejected by tidal 
torques through the outer Lagrange point, removing energy and angular 
momentum and forming a large tidal tail.  Its subsequent decompression 
may synthesize radioactive elements through the $r$-process 
\citep{Lattimer76,Freiburghaus99}, whose 
radioactive decay could power an optical transient \citep{Li98,Metzger10}.
For double NS binaries there are one or two such structures, depending 
on the mass ratio and equation of state (EoS) of the NSs \citep{Oechslin07}.  
Clearly, only one tail (or no tail) is formed in a BH-NS merger, and these 
 are typically a few thousand kilometers in size by the end of the 
disruption event and some of the fluid (as much as a few hundredths of 
a solar mass) is often gravitationally unbound.

The most efficient conversion of radioactive energy to radiation is provided by those 
isotopes with a decay timescale comparable to the radiative diffusion time 
through the ejecta.  In reality, there are likely to be a large number of 
nuclides with a very broad range of decay timescales \citep{Li98,Metzger10}. 
Current observational limits thus place interesting constraints on the 
nuclear evolution of this material, as well as the total mass ejected in 
these events. 

In this {\it Letter}, we present the results of multi-dimensional 
hydrodynamics simulations of neutron star mergers for various binary mass 
ratios.  The ejected material is then post-processed with a full nuclear 
network and a radiation transport code.  Our goal here is to investigate 
how radioactive $r$-process  elements synthesized in the tidal tails could 
power electromagnetic transients and how their properties may depend on 
the mass ratio of the merging binary.  We also discuss prospects for 
detection and implications for SGRBs.
\section{Tidal Tail Evolution}
\subsection{Ejection}

\begin{figure*}
\leavevmode
\begin{center}
\includegraphics[scale=0.83,angle=0]  {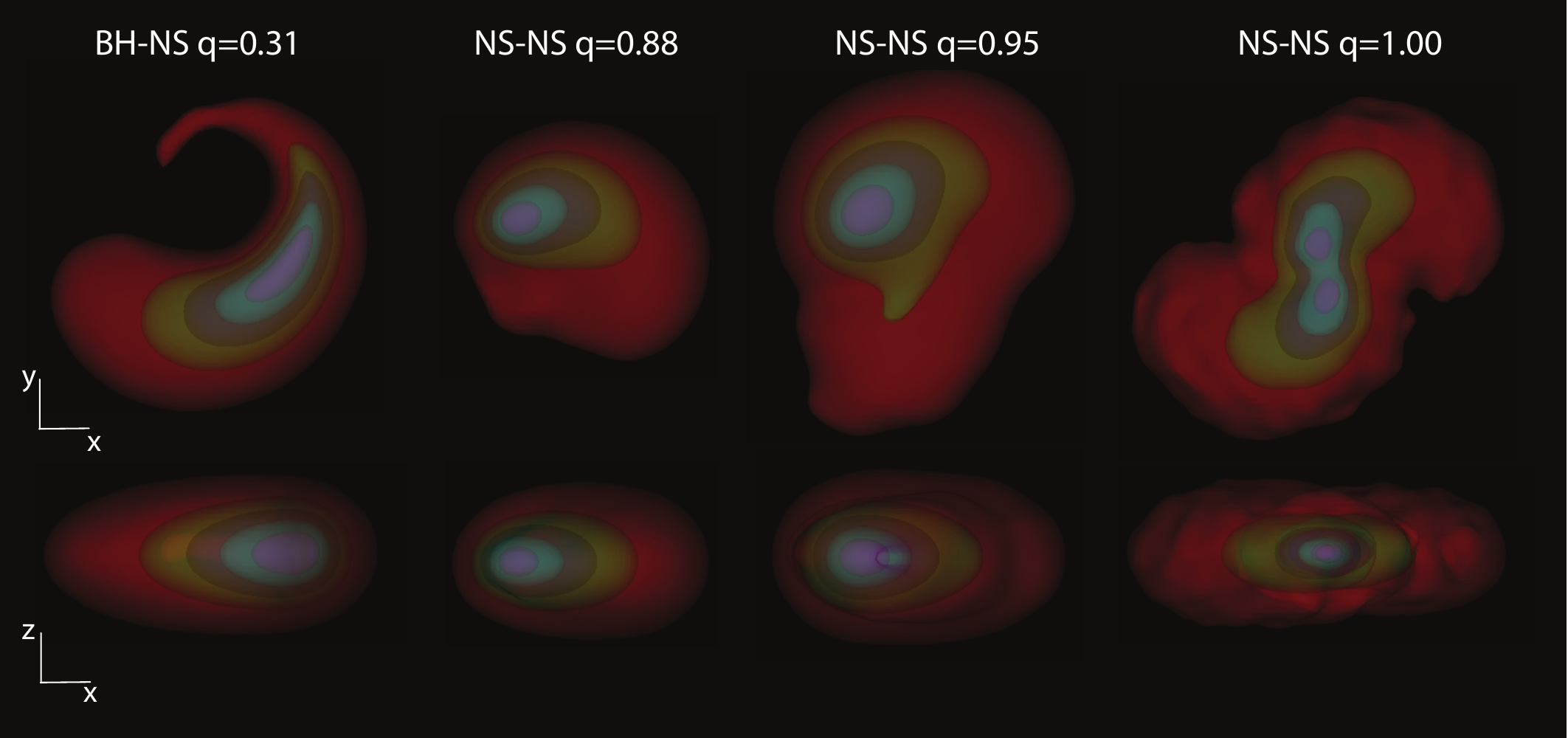}
\caption{Homologous structure of ejected material in the four models 
considered in this work (see table \ref{tab:models}) viewed face on 
(top) and edge on (bottom).  The density contours are not the same 
between models, we have instead chosen to plot the models at the same 
size scale.  The peak density contour (blue) is twenty times the minimum 
density contour (red).  The evolution from one to two tidal 
tails in the NS-NS mergers as the mass ratio approaches unity is clearly 
visible.}
\label{fig:homo_struct}
\end{center}
\end{figure*}   

To calculate the ejected mass and structure of the tidal tails, we 
employ a three dimensional smoothed particle hydrodynamics (SPH) method 
 to compute the evolution of the compact object merger \citep{Lee07,Lee10}.  
Due to its Lagrangian nature, SPH is perfectly suited to follow tidal 
disruption processes during which the corresponding geometry, densities, 
and time scales are changing violently \citep[c.f.][]{Rasio94,Lee00,Rosswog03}.  
The equation of state is a hybrid, where we have combined the cold 
Friedman-Pandharipande-Skyrme (FPS) nuclear EoS with an ideal thermal 
compoment as described by \citet{Shibata05}. Once the fluid in the 
tidal tails has become unbound, the full SPH calculations are stopped 
and the particles are followed on ballistic trajectories in the 
potential of the central compact remnant until their expansion becomes 
homologous.

To explore the possible range of outcomes depending on the physical 
parameters of the binary system, we have considered four models: 
three NS-NS mergers with mass ratios $q=M_{2}/M_{1}=1, 0.95, 0.88$ and one 
BH-NS merger with $q=0.31$. Around $~0.05 M_\odot$ of material is ejected in 
all four models. Relevant parameters resulting from the 
simulations are given in table \ref{tab:models}.  The density structure 
of the tails is shown in figure \ref{fig:homo_struct} at homology.  
Clearly, there is a progression from equal mass tails in the $q=1.0$ case 
to almost no secondary tail in the $q=0.88$ NS-NS case.  The relative mass
of the two tails is a function of the mass ratio and the nature of the 
merging objects, as well as the underlying EoS (which we have not explored 
in this work).  It is of interest to 
determine if different tail geometries are distinguishable in the optical 
signal from a compact object merger, as this would provide a significant 
constraint on the nature of the objects involved in addition to the expected
gravitational wave signal.

\begin{deluxetable*}{ccccccccc} 
\tablecolumns{7} 
\tablewidth{0pc} 
\tablecaption{Characteristics of simulations \label{tab:models}} 
\tablehead{ 
\colhead{Type}         &  \colhead{Mass Ratio}      & \colhead{Primary Mass}   & 
\colhead{Ejected Mass} &  \colhead{Ejecta Velocity} & \colhead{$t_{\rm h}$\tablenotemark{a}} & \colhead{$t_{\rm peak}$\tablenotemark{b}}    & 
\colhead{$t_{\rm dec}$\tablenotemark{c}} & \colhead{$L_{\rm peak}$\tablenotemark{d}}  \\ 
  &  & ($M_\odot$)  &  ($M_\odot$)  & ($c$) & (s) & (days) & (days) & (erg s$^{-1}$)}
\startdata 
NS-NS & 1.00  & 1.4 & 0.057 & 0.202  & 3.2 & 0.93 & 1.7 & $\sci{1.51}{42}$ \\
NS-NS & 0.95  & 1.4 & 0.047 & 0.200  & 2.0 & 0.93 & 1.6 & $\sci{1.19}{42}$ \\
NS-NS & 0.88  & 1.5 & 0.057 & 0.205  & 2.1 & 1.02 & 1.8 & $\sci{1.44}{42}$ \\
BH-NS & 0.31  & 5.4 & 0.060 & 0.248  & 3.9 & 0.93 & 1.7 & $\sci{1.64}{42}$ 
\enddata 
\tablenotetext{a}{Time to reach $1\%$ deviation from homologous evolution.}
\tablenotetext{b}{Time to reach peak bolometric luminosity after merger.}
\tablenotetext{c}{Light curve decay timescale.}
\tablenotetext{d}{Bolometric Luminosity.}
\end{deluxetable*}

\subsection{Nuclear Evolution}

\begin{figure}
\leavevmode
\includegraphics[scale=0.46]  {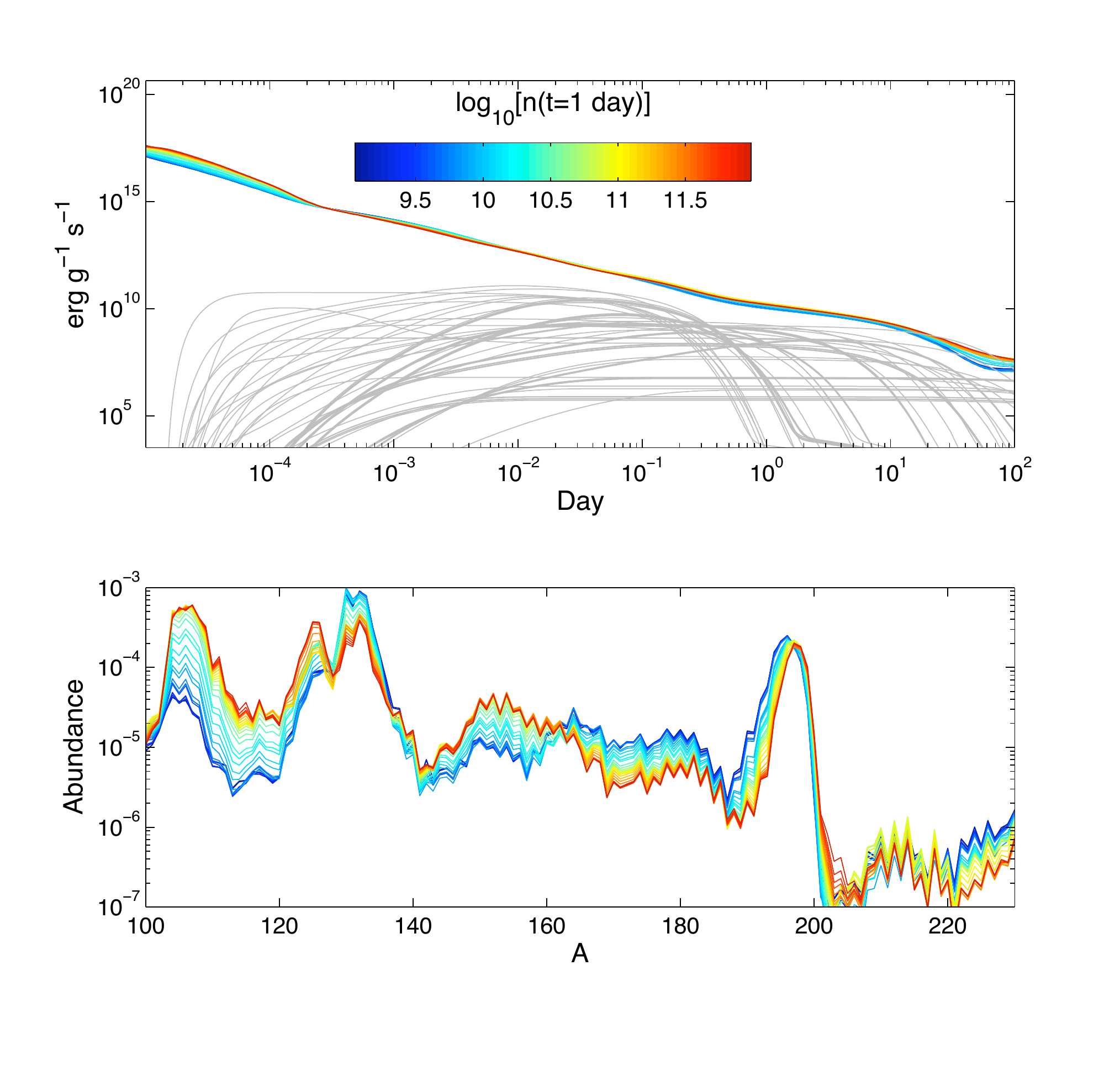}
\caption{Top: energy deposition rate from 
nuclear decay (including neutrino losses) as a function of time for 
various Lagrangian trajectories from the SPH simulations.  These are 
shown for the BH-NS merger, but are representative of the 
NS-NS mergers as well.  The trajectories are color coded by their 
density one day after the explosion.  The gray lines show the 
heating rate from single reactions that contribute significantly after 
0.1 days for a single trajectory.  Bottom: final 
abundances as a function of nuclear mass for the same trajectories.}
\label{fig:heating_rate}
\end{figure} 
  
We follow the evolution of the composition of the tails  using a 6312 isotope 
nuclear reaction network which extends past Uranium.  Density histories
of particles from the SPH simulations are employed.  The charged 
particle and neutron capture rates in the network up to At ($Z=85$) 
are taken from \cite{Rauscher00}.  Past At, the neutron capture rates 
of \cite{Panov10} are employed.  The network terminates at $Z=102$.  
Experimental values are taken for nuclear masses where available, 
elsewhere theoretical masses are taken from \cite{Moller03}. Neutron 
induced fission rates are taken from \cite{Panov10} and the simple 
approximation of \cite{Frankel47} is used to calculate spontaneous 
fission rates.  Fission barriers are taken from \cite{Mamdouh01}.
For our fission fragment distributions, we employ the empirical fits 
of \cite{Wahl02}.   

The nuclear network is evolved in time using a variant of the code 
{\verb XNet } \citep{Hix99}.  We have implemented the {\verb PARDISO }
sparse matrix solver \citep{Schenk08}, which makes calculations of 
large networks with implicitly coupled fission interactions feasible.  
The energy released by nuclear reactions is self-consistently added 
back to the material, similar to \cite{Freiburghaus99}.

The material ejected in the tail is from near the surface of the 
neutron star, making it challenging to determine the initial electron 
fraction and entropy of the material from our SPH simulations.  Therefore, 
we assume the initial electron fraction to be $Y_e = 0.2$ and start the 
calculation at $\rho = 10^{11} \textrm{g cm}^{-3}$ and $T_9 = 1$.  The 
initial composition is then determined by nuclear statistical equilibrium 
(NSE).  As was pointed out 
by \cite{Goriely05}, if tidally ejected material is un-shocked, the initial 
temperature of the material is likely much less than our assumed 
temperature.  To test the dependence on the initial temperature, we have 
run models going down to an initial temperature of $T_9 = 0.2$, and find 
that the nuclear heating rate is not substantially altered.

The total heating rate and final abundance distribution as a function 
of mass for a number of fluid elements are shown in figure 
\ref{fig:heating_rate}.  Similar to \cite{Metzger10}, we find that the 
late time heating rate is insensitive to the exact initial conditions and 
is statistical in nature, as was predicted by \cite{Li98}.  At one day, 
the top five beta-decays contributing to the heating rate are $^{125}$Sb, 
$^{126}$Sb, $^{132}$I, $^{127}$Te, and $^{197}$Pt.   

\subsection{Radiative Transfer}

In order to study the radiative transfer problem, we
employ the {\verb SEDONA } 3-D time-dependent LTE Monte Carlo
radiative transfer code which includes gamma-ray transfer and
spectropolarization \citep{Kasen06}.  The output of the hydrodynamic simulations 
is mapped into the radiative transfer code and the 
properties of the emergent radiation are calculated.  A global nuclear 
heating rate based on a fit to the nuclear network calculations is employed.
We approximately account for neutrino losses by assuming $75 \%$ of the 
nuclear network energy generation is deposited in the material \citep{Metzger10}.
Of the energy that is left, we assume $50\%$ is deposited as gamma rays
from decays while the other $50\%$ is deposited thermally.    

The opacity of r-process material at the relevant densities and 
temperatures is not well known.  The main contribution to the opacity 
is presumably due to millions of atomic lines, which are Doppler broadened 
by the high differential velocities in the ejecta. Unfortunately, complete 
atomic line lists for these high-Z species are not available.  Given the 
uncertainty, we assume in these calculations a constant gray opacity of 
$\kappa = 0.1~{\rm cm^2~g^{-1}}$ which is characteristic of the line 
expansion opacity from iron group elements \citep[e.g.][]{Kasen07}.  
This is in contrast to the work of \cite{Metzger10}, where oscillator 
strengths for pure Fe were assumed with ionization potentials from
Pb.  Considering that neither approach will yield accurate spectral
information, we feel that our simple grey opacity scheme is as 
reasonable an approximation as that used in \cite{Metzger10}.  

We directly calculated the spatial distribution of gamma-ray heating by 
following the transport of gamma-rays and determining the fraction of 
their energy thermalized by Compton scattering and photoelectric absorption.  
Since the dominant Compton opacity has only a weak wavelength dependence, 
the exact spectrum of gamma-ray emission from the radioactive source does 
not strongly affect the results.  We therefore simply assumed all gamma-rays 
were emitted at 1 MeV.  Before two days, the gamma-ray thermalization rate
is greater than 80\%.

\section{Detailed Properties of the Electromagnetic Counterparts} 

\begin{figure}
\leavevmode
\includegraphics[scale=0.6]  {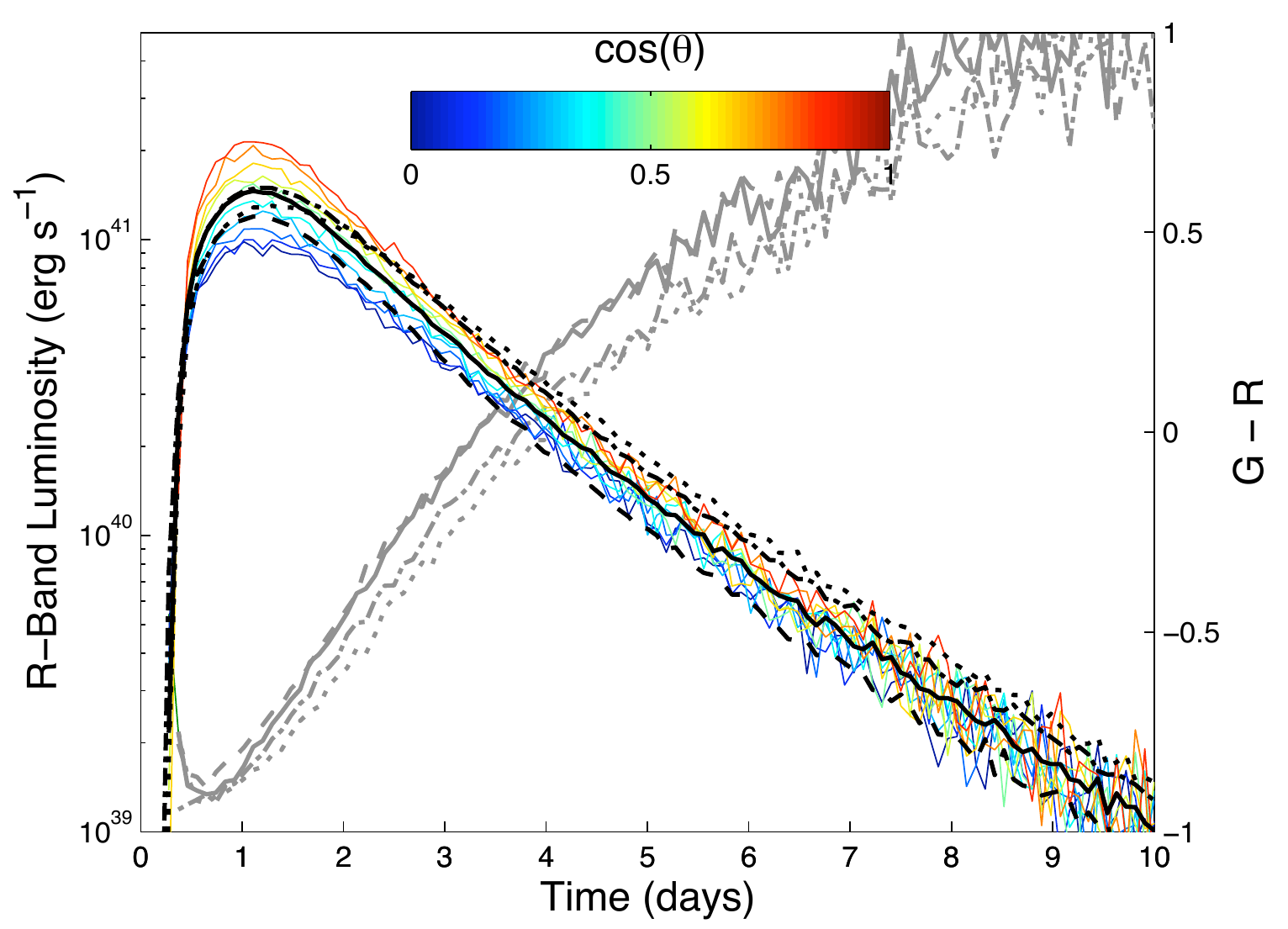}
\includegraphics[scale=0.6]  {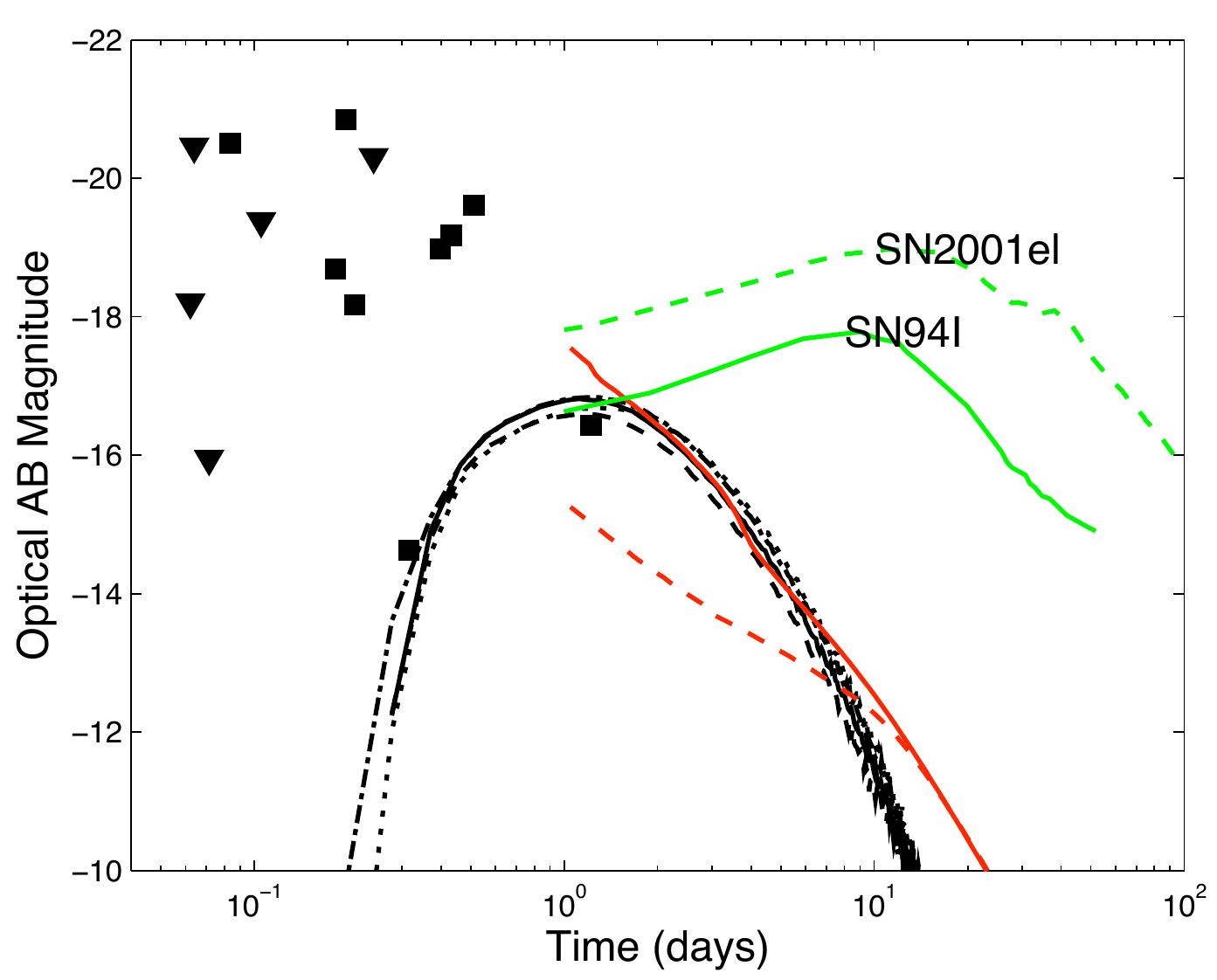}
\caption{Top: $r$-band luminosities for the three 
models described in table \ref{tab:models}.  The solid lines correspond to 
the $q=1.0$ NS-NS merger, the dashed lines to the $q=0.95$ 
NS-NS merger, the dashed lines to the $q=0.88$ 
NS-NS merger, and the dot-dashed lines to the BH-NS 
merger.  The black lines are the $r$-band luminosity for the models (left axis), averaged 
over all solid angles.  The gray lines are the $g-r$ color evolution for 
the models (right axis).
The colored lines give the luminosity as a function of polar angle 
(averaged over the phi direction) for the $q=1.0$ NS-NS merger.  Bottom:  $r$-band magnitudes
for the same models, along with observed SGRB afterglow absolute
magnitudes (filled squares) and upper limits (downward pointing 
triangles) from the compilation of \cite{Berger10} where red shifts 
were obtained.  All of the points correspond to separate events.  Synthetic 
SGRB afterglow light curves from 
\cite{VanEerten11} for a jet energy of $\sci{1}{48}\, \textrm{ergs}$ 
are also shown, where the solid (dashed) red line is for a jet opening angle of 0.2 (0.4) radian.  
The solid green curve is the optical light curve of SN94I, a type Ic 
supernova.  The dashed green curve is for SN2001el, a type Ia supernova.}
\label{fig:bolo_lum}
\end{figure}  

In the top panel of figure \ref{fig:bolo_lum}, the R-band light curves 
for the four models in table \ref{tab:models} are shown.  As would be 
expected from the simple models of \cite{Li98}, the peak luminosity
correlates with the total ejected mass of radioactive elements and the 
time of the peak scales inversely with mass and velocity.  Because the 
total mass ejected in these mergers is not very sensitive to $q$, the 
nature of the merger cannot easily be determined solely from the peak 
time or luminosity.  Additionally, the variation in peak luminosity 
with viewing angle within a single model is as large as the variation 
in the angle averaged peak luminosity between models.  This is also shown 
in figure \ref{fig:bolo_lum}.  This further complicates our ability 
to distinguish between different mass ratios and 
progenitor models based only on luminosities.\footnote{We also do not 
observe the non-smooth structures seen in the light curves of 
\cite{Metzger10}, which are due to their use of approximate non-grey
opacities.}    

Still, it may be possible to determine if one or two tails are present 
based on the color evolution of the light curves.  In NS-NS mergers 
which produce two tails, the luminosity of the transient will be given 
by the sum of the luminosities of the tails, each of which can be 
approximated as a \cite{Li98} expanding sphere.  We denote here the 
heavier and lighter tails with the subscript 1 and 2, respectively.
\cite{Li98} find that at late times, the evolution of the effective 
temperature is given by $T_{\rm eff} \approx \sci{4.7}{3} \, 
\textrm{K} \,  (M/0.01M_\odot)^{1/4} (c/v)^{3/4} 
(\textrm{\rm day}/t)^{3/4} (f/\sci{3}{-5})^{1/4}$.  The ratio of the 
effective temperatures in the tails at late times is then
\be
\frac{T_{\rm eff,1}}{T_{\rm eff,2}} \approx 
\left(\frac{M_1}{M_2}\right)^{1/4}
\left(\frac{v_2}{v_1}\right)^{3/4}.
\ee  
Significant variation in color from the single tail
case is then expected only for a considerable difference between 
velocities.  In this case, tail 2 will shift the total light blueward 
if it makes an important contribution to the total luminosity at any time.  
The time of peak luminosity for a single tail is given 
by $t_{\rm m} \approx 1 \, \textrm{day} \, (M/0.01M_\odot)^{1/2}
(3v/c)^{1/2}$.  If the velocity of the second tail is much 
lower, it contributes more to the total luminosity at late times 
and the light will be bluer compared to the more massive tail emitting
radiation alone.     

In our detailed models, the velocity difference between the tails is 
not significant enough compared to the mass difference between the
tails in either of the asymmetric NS-NS merger models for the tails 
to be easily discernible in their color evolution, as shown in figure 
\ref{fig:bolo_lum}.  As a result, there is no
significant distinguishing characteristic between ejecta 
geometries in their colors.  All four models show similar reddening 
as a function of time.

Without knowing the detailed line opacity of the R-process elements, it is difficult to predict the spectroscopic signatures of NS mergers.  We expect that high-Z atoms will have a very large number of lines in the optical, similar to and perhaps exceeding the number known for the iron group elements.  Because the ejecta velocities in the tidal tails are a factor of $\sim 2-3$, larger than ordinary supernovae, the absorption features in the photospheric phase ($\la 15$ days) should be very broad.  The likely outcome is a blending together of the lines into a relatively featureless continuum, not unlike the early time observations of broad line Type~Ic supernovae 
\citep[c.f.][]{Galama98}.

More detailed information about the ejecta geometry could be inferred from  spectroscopic observations taken at late times ($\ga 15$~days) when the ejecta has entered the nebular phase and emission features begin to appear.  Assuming relatively unblended lines (or line complexes) are present at these epochs, the spectral features should show double peaked emission profiles, at least for equatorial viewing angles.  The relative strength of the peaks could be used to estimate the relative masses of the two tails.

In brief, the inclusion of realistic ejecta geometries does not 
significantly alter the predictions made by simple spherical 
``tail'' models which include realistic nuclear physics.  This 
can be attributed to the fact that the ejecta does not deviate 
far  from a spherical geometry, the relative masses and velocities
of the two tails do no induce significant differences in the peak 
temperatures of the two tails, and we have assumed a gray opacity.  
Therefore, we have shown that the uncertainty does not lie in the 
structure of the ejecta, but in the total ejected mass, its velocity 
structure, and the opacity of pure $r$-process material.  The ejected 
mass and velocity are both dependent on the nuclear EoS
\citep{Lee00,Oechslin07} and the treatment of gravity \citep{Rosswog05}, both 
of which we have not studied in detail here.  Also, heating 
from $r$-process nucleosynthesis may effect the dynamics of the
material and the total ejected mass \citep{Metzger10b}.  The opacity of  
pure $r$-process material has not been studied at these low densities, so 
its relevance cannot be accurately assessed at this time.

\section{Implications for SGRBs, GW Observations, and the $r$-Process} 

\begin{deluxetable*}{ccccccccccccccccc} 
\tablecolumns{13} 
\tablewidth{0pc} 
\tablecaption{Detection Rates for Various Blind Transient Searches.  Properties of the surveys are assumed to be the same as 
those given in \cite{Strubbe09}.  \label{tab:rates}} 
\tablehead{ 
\colhead{} &\colhead{} &
\multicolumn{3}{c}{PTF}  &  \colhead{} & \multicolumn{3}{c}{Pan STARRS} &  \colhead{}  & 
\multicolumn{3}{c}{LSST} &  \colhead{}  &  \multicolumn{3}{c}{SASIR}\\
\cline{3-5} \cline{7-9} \cline{11-13} \cline{15-17}\\
\colhead{Model} & \colhead{} & \colhead{$R_{low}$}   & \colhead{$R_{re}$}    & \colhead{$R_{high}$} & \colhead{} 
  & \colhead{$R_{low}$}   & \colhead{$R_{re}$}    & \colhead{$R_{high}$} & \colhead{}
  & \colhead{$R_{low}$}   & \colhead{$R_{re}$}    & \colhead{$R_{high}$} & \colhead{}
  & \colhead{$R_{low}$}   & \colhead{$R_{re}$}    & \colhead{$R_{high}$}} \\
\startdata 
NS-NS &&$0.1$ &$10$  & $100$ &&$\sci{1.3}{-3}$ & $0.13$ & $1.3$ &&$\sci{2.3}{1}$ & $\sci{2.3}{3}$ & $\sci{2.3}{4}$ &&$1.2$ & $120$ & $\sci{1.2}{3}$\\
BH-NS &&$\sci{7.3}{-3}$ &$0.36$  & $12$ &&$\sci{8.4}{-5}$ & $\sci{4.2}{-3}$& $0.14$ &&$1.6$ & $79$ & $\sci{2.6}{3}$&&0.083& 4.1 & 140
\enddata 
\tablecomments{All rates are in detections per year.}
\end{deluxetable*}

If compact object mergers are indeed the progenitors of SGRBs, 
these optical transients could possibly be seen along side their more 
spectacular prompt gamma-ray display.  In the second 
panel of figure \ref{fig:bolo_lum}, we compare our synthetic optical 
light curves to optical afterglow observations and upper limits of 
SGRBs taken from \cite{Berger10}.  Clearly, the optical 
observations are all for on-axis SGRBs.  We also show two synthetic 
on-axis SGRB optical afterglow light curves taken from 
\cite{VanEerten11}, and for comparison, the light curves 
of two standard supernovae.  It seems that the optical observations can 
be reasonably explained by afterglow models, but, interestingly, they 
also do not rule out the possible contribution  of an $r$-process powered 
supernova.  It is also expected  that in a sizeable fraction of events, 
these $r$-process powered supernovae may dominate the optical light at 
timescales of a day. 

The current and next generation of surveys (such as PTF, SASIR, Pan STARRS and LSST) will be
sensitive to very subtle changes in flux, for sources with variability
timescales of more than a few days, across most of the sky.  In table 
\ref{tab:rates}, we show the detection rates expected for 
three transient surveys given the low, recommended, and high merger
rates given in \cite{Abadie10}, calculated assuming a cosmology 
given by \cite{Komatsu09} and a constant merger rate per volume.  The rates 
are corrected for the effects of a finite cadence.\footnote{Even a
cadence of 3 days reduces the detection rate by a factor of two compared
to when the effects of a finite cadence are left out.}   For the PTF and Pan STARRS, 
the detection rate is significantly lower than that expected for advanced 
LIGO whereas the inverse is true for SASIR and LSST.  This 
suggests that, if these large numbers of events are observed, optical 
observations will be more readily detectable than the associated gravity 
waves which could help focus gravitational wave searches and better 
constrain compact object merger rates.\footnote{We note that even 
if all compact mergers are associated with SGRBs, the $r$-process powered  
SNe will still dominate the off-axis optical afterglow.}  Additionally,
we find that LSST will potentially be able to detect these events out to 
$z>0.3$, giving  a handle on the evolution of the merger
rate with cosmic time.

It is clear from this and previous work 
\citep{Freiburghaus99,Goriely05} that large amounts of $r$-process 
material will be ejected in these events which may contribute 
significantly to the total  $r$-process budget of the galaxy, although 
there may be some problems explaining the low metallicity halo $r$-process 
abundance data within this scenario \citep{Argast04}.  Because the 
production of $r$-process elements is robust in the tails (i.e. the 
production of material with $A>120$ is insensitive to the initial 
conditions of the material), the major uncertainty in this model lies 
in the amount of mass ejected in these events and how this material
is spread through individual galaxies \citep{Zemp09,Kelley10}.  This is in 
contrast to the other possible site of $r$-process nucleosynthesis, 
neutrino driven winds, where the uncertainty lies in finding conditions 
that make an $r$-process \citep[c.f.][]{Roberts10}.  For a given amount 
of mass ejection, the total amount of $r$-process material in the Milky 
Way puts an upper limit on the merger rate \citep{Metzger10}.  Given these 
considerations, it seems likely that observations, or lack thereof 
of the transients produced in these mergers will give significant 
insight into the evolution of $r$-process material in the Universe.

\begin{acknowledgements}
Useful discussions with C. Fryer and S. Woosley 
are gratefully acknowledged. LR would like to thank Raph Hix 
for allowing us to use his nuclear network code and for many useful 
discussions concerning its use. We acknowledge support from an NNSA/DOE 
Stewardship Science Graduate Fellowship (LR) (DE-FC52-08NA28752), the 
University of California Office of the President (LR) (09-IR-07-117968-WOOS), 
the DOE SciDAC Program (DK) (DE-FC02-06ER41438),
CONACyT (WHL) (83254, 101958), and the  David and Lucille Packard Foundation 
(ERR) and the NSF (ERR) (AST-0847563).  Computing time was provided by 
ORNL through an INCITE award and by NERSC.
\end{acknowledgements}


\end{document}